\documentclass[journal]{IEEEtran}

\usepackage{amsmath}
\usepackage{nccmath}
\usepackage[utf8]{inputenc} 
\usepackage{epsfig,scalefnt,multirow}
\usepackage{url}
\usepackage{ifthen}
\usepackage{mathtools}
\usepackage{cite}
\usepackage{graphicx}
\usepackage{amssymb}
\usepackage{tabularx}
\usepackage{amsmath}
\usepackage{epstopdf}
\usepackage{algorithm}
\usepackage{algpseudocode}
\usepackage{algpascal}
\usepackage{cases}
\usepackage{stfloats}
\usepackage{xcolor}
\usepackage{tabularx}

  \def\cC{{\mathcal{C}}} 
\def\cE{{\mathcal{E}}}   
   
 \def\cN{{\mathcal{N}}}  
\def\cQ{{\mathcal{Q}}}

\def\diag{\mathop{\mathrm{diag}}}

\def\trace{\mathop{\mathrm{tr}}}

\def\Re{\mathop{\mathrm{Re}}}
\def\Im{\mathop{\mathrm{Im}}}

\def\b0{{\pmb{0}}} 

  \def\bg{{\mathbf{g}}} \def\bh{{\mathbf{h}}}
   
 \def\bn{{\mathbf{n}}}  
\def\bq{{\mathbf{q}}} \def\br{{\mathbf{r}}} \def\bs{{\mathbf{s}}} 
  \def\bw{{\mathbf{w}}} 
\def\by{{\mathbf{y}}}   

\def\bA{{\mathbf{A}}}  \def\bC{{\mathbf{C}}} 
   \def\bH{{\mathbf{H}}}
\def\bI{{\mathbf{I}}}  \def\bK{{\mathbf{K}}} 
\def\bM{{\mathbf{M}}} \def\bN{{\mathbf{N}}}  
 \def\bR{{\mathbf{R}}}  
  \def\bW{{\mathbf{W}}} \def\bX{{\mathbf{X}}}
\def\bY{{\mathbf{Y}}} 

\DeclarePairedDelimiter\norm{\lVert}{\rVert}

\begin{document}
	%
	\title{Channel Estimation for Spatially/Temporally Correlated Massive MIMO Systems \\ with One-Bit ADCs}
	%
	%
	%
	
	\author{Hwanjin~Kim	and~Junil~Choi
		\thanks{The authors are with the School of Electrical Engineering, Korea
			Advanced Institute of Science and Technology, Daejeon 34141, South Korea
			(e-mail: \{jin0903, junil\}@kaist.ac.kr).}
		\thanks{This work was partly supported by Institute for Information \& communications Technology Promotion(IITP) grant funded by the Korea government(MSIT) (No. 2016-0-00123, Development of Integer-Forcing MIMO Transceivers for 5G \& Beyond Mobile Communication Systems) and by the National Research Foundation (NRF) grant funded by the MSIT of the Korea government (2019R1C1C1003638).}
		\thanks{This paper was presented in part at the IEEE Global Communications Conference, 2018 \cite{2018Globecom}.}}

	\maketitle
	

\begin{abstract} 
This paper considers the channel estimation problem for massive multiple-input multiple-output (MIMO) systems that use one-bit analog-to-digital converters (ADCs). Previous channel estimation techniques for massive MIMO using one-bit ADCs are all based on single-shot estimation without exploiting the inherent temporal correlation in wireless channels. In this paper, we propose an adaptive channel estimation technique taking the spatial and temporal correlations into account for massive MIMO with one-bit ADCs. We first use the Bussgang decomposition to linearize the one-bit quantized received signals. Then, we adopt the Kalman filter to estimate the spatially and temporally correlated channels. Since the quantization noise is not Gaussian, we assume the effective noise as a Gaussian noise with the same statistics to apply the Kalman filtering. We also implement the truncated polynomial expansion-based low complexity channel estimator with negligible performance loss. Numerical results reveal that the proposed channel estimators can improve the estimation accuracy significantly by using the spatial and temporal correlations of channels.
\end{abstract}






\section{Introduction}\label{sec:introd}
Massive multiple-input multiple-output (MIMO) systems are one of the promising techniques for next generation wireless communication systems \cite{Marzetta06,Rusek13,Hoydis13,Larsson14}. By using a large number of antennas at base stations (BSs), it is possible to support multiple users simultaneously to boost network throughput and improve the energy efficiency by beamforming techniques \cite{Hoydis13}. Due to the large number of antennas at the BS, high implementation cost and power consumption could be major problems for implementing massive MIMO in practice.

Using low-resolution analog-to-digital converters (ADCs) is an effective way of mitigating the power consumption problem in massive MIMO systems because the ADC power consumption exponentially decreases as its resolution level \cite{Walden99}. However, symbol detection and channel estimation in massive MIMO systems with low-resolution ADCs become difficult tasks because the quantization process using low-resolution ADCs becomes highly nonlinear. 
Recent works have revealed that it is possible to implement practical symbol detectors and channel estimators for massive MIMO even with low-resolution ADCs. For the symbol detection, a massive spatial modulation MIMO approach based on sum-product-algorithm was developed in \cite{Wang15}, a convex optimization based multiuser detection for massive MIMO with low-resolution ADC was considered in \cite{Wang14}, a mixed-ADC massive MIMO detector was proposed in \cite{Zhang16}, and a blind detection technique was developed by exploiting supervised learning \cite{Jeon17}. Also, an iterative detection and decoding scheme based on the message passing algorithm and low resolution aware (LRA) minimum mean square error (MMSE) receive filter was presented in \cite{Shao18}, a low complexity maximum likelihood detection (MLD) algorithm called one-bit-sphere-decoding was developed in \cite{Jeon18}, and a successive cancellation soft-output detector by exploiting a previous decoded message was proposed in \cite{Cho19}.

For the channel estimation, a near maximum likelihood channel estimator based on the convex optimization was developed in \cite{Choi16}, and a Bayes-optimal joint channel and data estimator was proposed in \cite{Wen16}. To reduce the complexity, the generalized approximate message passing algorithm was applied in \cite{Jmo17}, and the hybrid architectures were considered in \cite{Jmo171}. Moreover, an oversampling based LRA-MMSE channel estimator that exploits the correlation of filtered noise for a given channel was proposed in \cite{Shao19}. However, up to the authors' knowledge, the previous channel estimators with low-resolution ADCs have not considered the temporal correlation in channels, which is inherent in communication channels.

In this paper, we develop a channel estimator taking both spatial and temporal correlations into consideration for massive MIMO systems with one-bit ADCs. We first discuss how to estimate the spatial correlation matrix for the channel estimation. Then, we reformulate the non-linear one-bit quantizer to the linear operator based on the Bussgang decomposition\cite{Bussgang52}. To exploit both the spatial and temporal correlations, we implement the Kalman filter-based (KFB) estimator \cite{Kalman} assuming the statistically equivalent quantization noise after the Bussgang decomposition follows a Gaussian distribution with the same mean and covariance matrix. The numerical results demonstrate that the normalized mean square error (NMSE) of the proposed KFB estimator decreases as the time slot increases. Moreover, as channels are more correlated with space and time, it is possible to track the channels more accurately. 
To reduce the complexity of KFB estimator, which comes from the large size matrix inversion, we also exploit a truncated polynomial expansion (TPE) approximation for the matrix inversion in the Kalman gain matrix. We analytically show that, with some moderate assumptions, the NMSE of the TPE-based estimator also keeps decreasing with the time slots. The numerical results show that the low-complexity TPE-based estimator gives approximately the same performance as the KFB estimator even with low approximation orders.

The rest of the paper is organized as follows. In Section \ref{sec:model}, we explain a system model with one-bit ADCs. In Section \ref{sec:analysis}, we first discuss how to estimate the spatial correlation matrix. Then, we review the single-shot channel estimator based on the Bussgang decomposition \cite{Li17}. After, we explain our proposed successive channel estimator based on the Bussgang decomposition and the Kalman filter. We also propose the low-complexity TPE-based channel estimator and analyze the complexities of competing estimators. After explaining the data transmission with one-bit ADCs in Section \ref{sec:datatransmission}, we  evaluate numerical results in Section \ref{sec:numericalresult}. Finally, we conclude the paper in Section \ref{sec:conclusions}. 

\textbf{Notation:} Lower and upper boldface letters represent column vectors and matrices. $\bA^{\mathrm{T}}$, $\bA^{*}$,  $\bA^{\mathrm{H}}$, and $\bA^{\dagger}$ denotes the transpose, conjugate, conjugate transpose, and pseudo inverse of the matrix $\bA$. $\mathbb{E}\{\cdot\}$ represents the expectation, and $\Re\{\cdot\}$, $\Im\{\cdot\}$ denote the real part and imaginary part of the variable. $\boldsymbol{0}_m$ is used for the $m\times1$ all zero vector, and $\bI_m$ denotes the $m \times m$ identity matrix. $\otimes$ denotes the Kronecker product. $\diag(\cdot)$ returns the diagonal matrix. $\mathrm{vec}(\cdot)$ denotes the columnwise vectorization. ${\mathbb{C}}^{m \times n}$ and ${\mathbb{R}}^{m \times n}$ represent the set of all $m \times n$ complex and real matrices. $|{\cdot}|$ denotes the amplitude of the scalar, and $\norm{\cdot}$ represents the $\ell_2$-norm of the vector. $\cC \cN(m,\sigma^2)$ denotes the complex normal distribution with mean $m$ and variance $\sigma^2$. $\trace(\cdot)$ represents the trace operator. $\mathcal{O}$ denotes the Big-O notation.

\section{System Model}\label{sec:model}
\begin{figure}[t]
	\centering
	\includegraphics[width=8.8 cm]{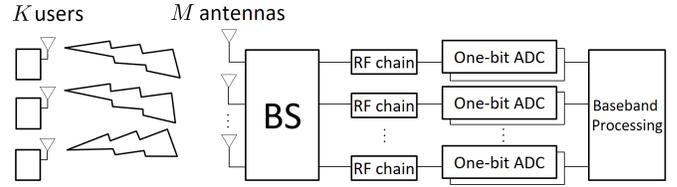}%
	\caption{Massive MIMO systems with $M$ BS antennas and $K$ single-antenna users. Each RF chain is equipped with two one-bit ADCs for the in-phase and quadrature component, respectively.}\label{fig:system}
\end{figure}

As in Fig. \ref{fig:system}, we assume a single-cell massive MIMO system with $M$ BS antennas and $K$ single-antenna users with $M\gg K$. 
Each BS antenna is connected to two one-bit ADCs; one for the in-phase component and the other for the quadrature component of received signals. We consider the block-fading channel with the coherence time of $T$. The received signal at the $i$-th fading block is given by
\begin{align}
\by_i=\sqrt{\rho}\bH_i\bs_i+\bn_i, \label {model}
\end{align}
where $\rho$  is the signal-to-noise ratio (SNR), $\bH_i=[\bh_{i,1},\bh_{i,2},...,\bh_{i,K}]\in {\mathbb{C}}^{M\times K}$ is the channel matrix, $\bh_{i,k}$ is the channel between the BS and the $k$-th user in the $i$-th fading block, $\bs_i$ is the transmit signal, and $\bn_i\sim \cC \cN(\boldsymbol{0}_{M},\bI_{M})$ is the complex Gaussian noise. We consider the spatially and temporally correlated channels by assuming $\bh_{i,k}$ follows the first-order Gauss-Markov process, 
\begin{align}
\bh_{0,k}&=\bR_k^{\frac{1}{2}}\bg_{0,k},\notag\\
\bh_{i,k}&=\eta_k\bh_{i-1,k}+\sqrt{1-\eta_k^2}\bR_k^{\frac{1}{2}}\bg_{i,k}, \quad i\geq1, \label{channel_model}
\end{align}
where $\eta_k$ is the temporal correlation coefficient, $\bR_k={\mathbb{E}}\{\bh_{i,k}\bh_{i,k}^\mathrm{H}\}$ is the spatial correlation matrix, and $\bg_{i,k}\sim \cC \cN(\boldsymbol{0}_{M},\bI_{M})$ is the innovation process of the $k$-th user in the $i$-th fading block. Note that $\eta_k$ and $\bR_k$ do not have the time index $i$ since both are long-term statistics that are static for multiple coherence blocks.

Although other models are also possible, to have concrete analyses, we adopt the exponential model for the spatial correlation matrix $\bR_k$,
\begin{align}
\bR_k=\begin{bmatrix} 
1 &r_k &\cdots &r_k^{M-1}~\\
r_k^* &1  &\cdots  &\vdots\\
\vdots &\vdots &\ddots &\vdots\\
r_k^{*(M-1)} &\cdots &\cdots &1~\\
\end{bmatrix},
\end{align}
where $r_k=re^{j\theta_k}$ satisfying $0\leq r<1$ and $0\leq\theta_k<2\pi$. We assume all users experience the same spatial correlation coefficient $r$ since it is dominated by the BS antenna spacing while each user has an indifferent phase $\theta_k$ since it is more related to the user position \cite{Clerckx08}.

The quantized signal by the one-bit ADCs is
\begin{align}
\br_i=\mathcal{Q}(\by_i)=\mathcal{Q}( \sqrt{\rho}\bH_i\bs_i+\bn_i), \label {quatized signal}
\end{align}
where $\mathcal{Q}(\cdot)$ is the element-wise one-bit quantization operator, i.e., ${\mathcal{Q}}(\cdot)=\frac{1}{\sqrt{2}}(\mathrm{sign}({\Re}\{\cdot\})+j~\mathrm{sign}({\Im}\{\cdot\})).$

\section{Channel Estimation with One-Bit ADCs}\label{sec:analysis}	
In this section, we first discuss how to estimate the spatial correlation matrix. Then, we explain the Bussgang linear minimum mean square error (BLMMSE) estimator, which is the baseline of the proposed estimator. The BLMMSE estimator is a single-shot channel estimator based on the Bussgang decomposition without exploiting any temporal correlation \cite{Li17}. Then, we propose the KFB estimator, which is a successive channel estimator, for massive MIMO with one-bit ADCs exploiting the temporal correlation. Also, we propose the low-complexity TPE-based estimator to reduce the complexity of the proposed KFB estimator.  

\subsection{Spatial correlation matrix estimation}
In this subsection, we discuss how to estimate the spatial correlation matrix since all the channel estimators in this paper exploit the spatial correlation of channel. We omit the user index $k$ since the BS can estimate the spatial correlation of each user separately.

When the BS does not have any prior channel information, it can use the least square (LS) estimate of the quantized signal $\br_i$, which is given by
\begin{align}
{\bh^{\textrm{LS}}_i=\boldsymbol{\Phi}_i^\dagger\br_i,}
\end{align}
where $\boldsymbol{\Phi}_i$ is the pilot matrix. The LS estimator for one-bit quantized signal performs well when the number of antennas at the BS is large, as shown in \cite{Choi141}. The BS then can obtain a sampled spatial correlation matrix as 
\begin{align}
\hat{\bR}=\frac{1}{N_s}\sum_{n=1}^{N_s}\bh_{n}^{\textrm{LS}}\left(\bh_{n}^{\textrm{LS}}\right)^\mathrm{H},
\end{align}
where $N_s$ is the number of samples. We evaluate the performance loss by using the sampled correlation matrix in Fig. 3 in Section 5. After this subsection, we assume that the true spatial correlation matrices and the temporal correlation coefficients of all users are known to the BS to derive analytical results.

\subsection{BLMMSE estimator}\label{single}
In this subsection, we omit the time slot index $i$ since the single-shot channel estimator does not use any temporal correlation. To estimate the channel at the BS, $K$ users transmit the length $\tau$ pilot sequences to the BS,
\begin{align}
{\mathbf{Y}}=\sqrt{\rho}{\mathbf{H}}{\boldsymbol{\Phi}}^\mathrm{T}+{\mathbf{N}},
\end{align}
where $\bY\in{\mathbb{C}}^{M\times\tau}$ is the received signal, ${\boldsymbol{\Phi}} \in {\mathbb{C}}^{\tau \times K}$ is the pilot matrix, and $\mathbf{N}=[\bn_1,\bn_2,...,\bn_{\tau}]\in{\mathbb{C}}^{M\times\tau}$ is the complex Gaussian noise.
We assume that the pilot sequences are column-wise orthogonal, i.e., ${\boldsymbol{\Phi}}^\mathrm{T}{\boldsymbol{\Phi}}^*=\tau \bI_K$, and all the elements of the pilot matrix have the same magnitude. For the sake of simplicity, the receive signal is vectorized as
\begin{align}
{\mathrm{vec}}(\bY)=\underline{\by}=\bar{\boldsymbol{\Phi}}\underline{\bh}+\underline{\bn}, \label{y}
\end{align}
where $\bar{\boldsymbol{\Phi}}=(\boldsymbol{\Phi}\otimes\sqrt{\rho}\bI_M)$, $\underline{\bh}=\mathrm{vec}(\bH)$,  and $\underline{\bn} = \mathrm{vec}(\bN)$. The quantized signal by one-bit ADCs is 
\begin{align}
\underline{\br}={\mathcal{Q}}(\underline{\by}).\label{quantized}
\end{align}

Assuming independent spatial correlations across the users, the aggregated spatial correlation matrix $\bR={\mathbb{E}}\{\underline{\bh} \underline{\bh}^\mathrm{H}\} $ is given by
\begin{align}
\bR=\begin{bmatrix} 
\bR_1 &\cdots &0 &0\\
\vdots &\bR_2 &\cdots &0\\
0 &\vdots &\ddots  &\vdots\\
0 &0 &\cdots &\bR_K \label{R_k}
\end{bmatrix}.
\end{align} 

The Bussgang decomposition of quantized signal is given by
\begin{align}
\underline{\br}={\mathcal{Q}}(\underline\by)=\bA\underline\by+\bq, \label{Quantizer}
\end{align}
where $\bA$ denotes the linear operator and $\bq$ represents the statistically equivalent quantization noise. The linear operator $\bA$ is obtained from \cite{Li17},
\begin{align}
\bA&=\sqrt{\frac{2}{\pi}}\diag(\bC_{\underline\by})^{-\frac{1}{2}}\notag \\
&=\sqrt{\frac{2}{\pi}}\diag\left(\bar{\boldsymbol{\Phi}}{\bR}\bar{\boldsymbol{\Phi}}^\mathrm{H}+\bI_{M\tau}\right)^{-\frac{1}{2}}\notag\\
&\stackrel{(a)}{=}\sqrt{\frac{2}{\pi}}\sqrt{\frac{1}{K\rho+1}}\bI_{M\tau},\label{A}
\end{align}
where $\bC_{\underline\by}$ is the auto-covariance matrix of the received signal.
In (\ref{A}), $(a)$ is derived in Appendix A.
Substituting (\ref{y}) into (\ref{Quantizer}), $\underline\br$ is represented as
\begin{align}
\underline\br={\mathcal{Q}}(\underline\by)=\tilde{{\boldsymbol{\Phi}}}\underline{\bh}+\tilde{\underline\bn},
\end{align}
where $\tilde{\boldsymbol{\Phi}}=\bA\bar{\boldsymbol{\Phi}}\in{\mathbb{C}}^{M\tau\times{MK}}$ and $ \tilde{\underline\bn}=\bA\underline{\bn}+\bq\in{\mathbb{C}}^{M\tau\times 1}$.

After adopting the Bussgang decomposition, a linear MMSE estimator, which is denoted as the BLMMSE channel estimator \cite{Li17}, is given as
\begin{align}
\underline{\hat{\bh}}^{\mathrm{BLM}}=\bC_{\underline{\bh}}\tilde{\boldsymbol{\Phi}}^\mathrm{H}\bC_{\underline\br}^{-1}\underline\br,\label{BLM}
\end{align}
where $\bC_{\underline\bh}$ is the auto-covariance matrix of the channel $\underline\bh$, and $\bC_{\underline\br}$ is the auto-covariance matrix of the quantized signal $\underline\br$.
In (\ref{BLM}), $\bC_{\underline\br}$ is obtained by the arcsin law \cite{arcsin},
\begin{align}
\bC_{\underline\br}&=\frac{2}{\pi}\Big(\arcsin\Big(\Sigma_{\underline\by}^{-1/2}\Re\{\bC_{\underline\by}\}\Sigma_{\underline\by}^{-1/2}\Big)\notag\\&+j\arcsin\Big(\Sigma_{\underline\by}^{-1/2}\Im\{\bC_{\underline\by}\}\Sigma_{\underline\by}^{-1/2}\Big)\Big),\label{arcsinlaw}
\end{align}
where $\Sigma_{\underline\by}=\diag(\bC_{\underline\by})$.


\subsection{Proposed KFB estimator}{\label{Successive}}
Although effective, the BLMMSE estimator does not exploit any inherent temporal correlation in wireless channels. We now propose a simple, yet effective channel estimator based on the Bussgang decomposition and the Kalman filtering. We recover the time slot index $i$ to explicitly use the temporal correlation. We first reformulate the channel model in (\ref{channel_model}) by vectorization,
\begin{align}
\underline{\bh}_0&={\bR}^{\frac{1}{2}}\underline{\bg}_0,\notag \\
\underline{\bh}_i&={\boldsymbol{\eta}}\underline{\bh}_{i-1}+{\boldsymbol{\zeta}} {\bR}^{\frac{1}{2}}\underline{\bg}_i, ~~~ i\geq1, \label{channel}
\end{align}
where $\underline{\bg}_i$ is the vectorized innovation process, which is expressed as
\begin{align}
\underline{\bg}_i=\left[\bg_{i,1}^\mathrm{T},\bg_{i,2}^\mathrm{T},...,\bg_{i,K}^\mathrm{T}\right]^\mathrm{T},~~~ i\geq0.
\end{align}
The temporal correlation matrices ${\boldsymbol{\eta}}$ and ${\boldsymbol{\zeta}}$ in (\ref{channel}) are given by the Kronecker product,
\begin{align}
{\boldsymbol{\eta}}&=\diag({\eta}_1,{\eta}_2,...,{\eta}_K) \otimes \bI_M,\notag \\
{\boldsymbol{\zeta}}&=\diag({\zeta}_1,{\zeta}_2,...,{\zeta}_K) \otimes \bI_M,
\end{align}
where $\eta_k$ denotes the $k$-th user temporal correlation coefficient and $\zeta_k=\sqrt{1-\eta_k^2}$. 

Following the same steps as in Section \ref{single}, the one-bit quantized signal can be represented using the Bussgang decomposition as
\begin{align}
\underline{\br}_i&={\mathcal{Q}}(\underline{\by}_i),\label{quantized_i}\\
&=\bA_i\underline{\by}_i+\bq_i, \label{Quantizer_i}\\
&=\tilde{{\boldsymbol{\Phi}}}_i\underline{\bh}_i+\tilde{\underline{\bn}}_i, \label{receive_signal}
\end{align}
where $\bA_i$ is the linear operator, $\bq_i$ is the statistically equivalent quantization noise,   $\tilde{\boldsymbol{\Phi}}_i=\bA_i\bar{\boldsymbol{\Phi}}_i\in{\mathbb{C}}^{M\tau\times {MK}}$, and $ \tilde{\underline\bn}_i=\bA_i\underline{\bn}_i+\bq_i\in{\mathbb{C}}^{M\tau\times 1}$.

\begin{algorithm}[t]
	\begin{algorithmic}[1]
		\caption{Kalman Filter-Based Channel Estimator}
		\State Initialization:
		\begin{align*}
		\hat{\underline{\bh}}_{0|0}=\boldsymbol{0}_{MK},~
		{\bM}_{0|0}={\bR}={\mathbb{E}}\left\{\underline{\bh}_0\underline{\bh}_0^\mathrm{H}\right\}
		\end{align*}
		\State Prediction:
		\begin{align*}
		\underline{\hat{\bh}}_{i|i-1}={\boldsymbol{\eta}}\underline{\hat{\bh}}_{i-1|i-1}
		\end{align*}
		\State Minimum prediction MSE matrix ($MK\times MK$):
		\begin{align*}
		\bM_{i|i-1}={\boldsymbol{\eta}}\bM_{i-1|i-1}{\boldsymbol{\eta}}^\mathrm{H}+{\boldsymbol{\zeta}}{\bR}{\boldsymbol{\zeta}}^\mathrm{H}
		\end{align*}
		\State Kalman gain matrix ($MK\times M\tau$):
		\begin{align*}
		\bK_i=\bM_{i|i-1}\tilde{\boldsymbol{\Phi}}_i^\mathrm{H}\left(\bC_{\tilde{\underline{\bn}}_i}+\tilde{\boldsymbol{\Phi}}_i\bM_{i|i-1}\tilde{\boldsymbol{\Phi}}_i^\mathrm{H}\right)^{-1}
		\end{align*}
		\State Correction:
		\begin{align*}
		\hat{\underline{\bh}}_{i|i}=\hat{\underline{\bh}}_{i|i-1}+\bK_i\left({\underline{\br}_i}-\tilde{\boldsymbol{\Phi}}_i\hat{\underline{\bh}}_{i|i-1}\right)   
		\end{align*}
		\State Minimum MSE matrix ($MK\times MK$):
		\begin{align*}
		\bM_{i|i}=\left(\bI_{MK}-\bK_i\tilde{\boldsymbol{\Phi}}_i\right)\bM_{i|i-1}
		\end{align*}
	\end{algorithmic}
\end{algorithm} 
The Kalman filter guarantees the optimality when the noise is Gaussian distributed \cite{Kalman}; however, the effective noise $\tilde{\underline\bn}_i$ in (\ref{receive_signal}) is not Gaussian because of the one-bit quantization noise $\bq_i$. Although the noise is not Gaussian, it is still possible to apply the Kalman filter using the same covariance matrix $\bC_{\tilde{\underline\bn}_i}$. The proposed KFB channel estimator is summarized in Algorithm 1. 


\textbf{Remark 1: } Assuming the effective noise is Gaussian distributed may result in inaccurate channel estimation. This effect becomes more dominant as SNR increases, which is shown in Fig. 4 in Section \ref{sec:numericalresult}. In the high SNR regime, the noise $\underline{\bn}_i$ in (\ref{model}) becomes negligible, and the effective noise $\tilde{\underline\bn}_i$ in (\ref{receive_signal}) is dominated by the quantization noise $\bq_i$, which would severely violate the Gaussian assumption of $\tilde{\underline{\bn}}_i$. In the low SNR regime, however, the effective noise $\tilde{\underline\bn}_i$ is more like Gaussian, and the proposed KFB estimator is nearly optimal.

\subsection{Low-complexity TPE-based estimator}{\label{Reduction}}
The BLMMSE estimator is a single-shot estimator, which returns a new channel estimate while the KFB estimator is a successive channel estimator, which tracks the channel based on a previous channel estimate at each time slot. Thus, the complexity of both channel estimators is the same at each time slot. 

The matrix inversion has the most dominant computation complexity among matrix operations. The large channel dimensions in massive MIMO systems even exacerbate the complexity of matrix inversion. Therefore,  when comparing the complexity of algorithms, we only consider the complexity of the matrix inversion. To reduce the complexity of KFB estimator, the truncated polynomial expansion \cite{Shariati14} can be used to approximate the matrix inversion at the Kalman gain matrix $\bK_i$ in Step 4 of Algorithm 1. 

The $L^\mathrm{th}$-order TPE approximation of the inversion of $N\times N$ matrix $\bX$ is expressed as
\begin{align}
\bX^{-1}\approx \alpha \sum_{l=0}^L (\bI-\alpha\bX)^l \label{TPE}.
\end{align}
In (\ref{TPE}), $\alpha$ is the convergence coefficient, which can be set as $0<\alpha<\frac{2}{\max_n \lambda_n(\bX)}$ where $\lambda_n(\bX)$ is the $n$-th eigenvalue of the matrix $\bX$ \cite{Shariati14}. 

The complexity of TPE approximation in (\ref{TPE}) is $\mathcal{O}(LN^2)$ since it has only the matrix multiplication with the $L^{\mathrm{th}}$-order. This is a large complexity reduction as compared to $\mathcal{O}(N^3)$ for the complexity of the $N\times N$ matrix inversion when $L$ is much smaller than $N$. In Table I, we summarize the complexity of three competing estimators. The TPE-based estimator has much lower computational complexity than the other estimators because $L\ll M\tau$ in practice.  
\begin{table}[tbp]{\label{table1}}
	\renewcommand{\arraystretch}{1.3}
	\caption{Computational complexity of BLMMSE estimator, KFB estimator, and TPE-based estimator. $M$: number of antennas, $\tau$: pilot symbol length, $L$: approximation order}
	\label{tab:table_example}
	\centering
	\begin{tabularx}{\columnwidth}{l  l}
		\hline
		\text{Channel estimator}~~~~~~~ & \text{Computational complexity}\\
		\hline
		BLMMSE estimator~~~~~~~ & $\mathcal{O}(M^3\tau^3)$\\
		KFB estimator~~~~~~~ & $\mathcal{O}(M^3\tau^3)$\\
		TPE-based estimator~~~~~~~ & $\mathcal{O}(LM^2\tau^2)$\\
		\hline
	\end{tabularx}
\end{table}

To verify the effectiveness of TPE approximation, we evaluate the minimum NMSE of TPE-based estimator. For a tractable analysis, we assume ${\bR}=\bI_{MK}$ and $\tau=K$ as in \cite{Li17}, which results in $\bC_{\underline\br}=\bI_{MK}$ in (\ref{arcsinlaw}) since $\bC_{\underline\by}=(K\rho+1)\bI_{MK}$. Then the NMSE of BLMMSE estimator in \cite{Li17}, which is a performance baseline of the proposed estimators, is represented as 
\begin{align}
\mathrm{NMSE}_{\mathrm{BLM}}&=
\frac{1}{MK}{\mathbb{E}}\left\{ \left\|\hat{\underline{\bh}}^{\mathrm{BLM}}-\underline{\bh}\right\|_2^2\right\}\notag\\
&=1-\frac{2}{\pi}\frac{K\rho}{K\rho+1}\notag\\
&=1-\beta, \label{BLMMSE_NMSE2}
\end{align}
where 
$\beta=\frac{2}{\pi}\frac{K\rho}{K\rho+1}$.

To derive the NMSE of TPE-based estimator, we first expand the covariance matrix of $\bq_i$ as 
\begin{align}
{\bC_{\bq_{i}}}&\stackrel{~}{=}\bC_{\underline{\br}_{i}}-\bA_{i}\bC_{\underline{\by}_{i}}\bA_{i}^\mathrm{H} \notag\\
\notag&\stackrel{(a)}{=}\bC_{\underline{\br}_{i}}-\frac{2}{\pi}(\bI_{MK})\\
\notag&\stackrel{(b)}{=}\frac{2}{\pi}(\text{arcsin}(\bX_i)+j \text{arcsin}(\bY_i))-\frac{2}{\pi}(\bI_{MK})\\
&\stackrel{(c)}{=}\left(1-\frac{2}{\pi}\right)\bI_{MK},\label{covariancematrixqi}
\end{align}
where we define
\begin{align}
\bX_i=\Sigma_{\underline\by_{i}}^{-1/2}\Re\{\bC_{\underline\by_{i}}\}\Sigma_{\underline\by_{i}}^{-1/2},\notag\\
\bY_i=\Sigma_{\underline{\by}_{i}}^{-1/2}\Im\{\bC_{\underline\by_{i}}\}\Sigma_{\underline\by_{i}}^{-1/2}.\label{X_i}
\end{align}
In (\ref{covariancematrixqi}), $(a)$ comes from $\bA_i=\sqrt{\frac{2}{\pi}}\sqrt{\frac{1}{K\rho+1}}\bI_{MK}$ and $\bC_{\underline\by_i}=(K\rho+1)\bI_{MK}$, $(b)$ is from the arcsin law in \cite{arcsin}, and $(c)$ is derived by substituting $\bC_{\underline\by_i}=(K\rho+1)\bI_{MK}$ into (\ref{X_i}).

The first-order TPE approximation of matrix inversion in Kalman gain matrix is given by
\begin{align}
\bX^{-1}\approx \alpha (\bI+(\bI-\alpha\bX)).
\end{align}
Thus, the Kalman gain matrix is approximated as
\begin{align}
\bK_i&=\bM_{i|i-1}\tilde{\boldsymbol{\Phi}}_i^\mathrm{H}\left(\bC_{\tilde{\underline{\bn}}_i}+\tilde{\boldsymbol{\Phi}}_i\bM_{i|i-1}\tilde{\boldsymbol{\Phi}}_i^\mathrm{H}\right)^{-1}\notag \\
&\approx \bM_{i|i-1}\tilde{\boldsymbol{\Phi}}_i^\mathrm{H}\left(2\alpha\bI_{MK}-\alpha^2\left(\bC_{\tilde{\underline{\bn}}_i}+\tilde{\boldsymbol{\Phi}}_i\bM_{i|i-1}\tilde{\boldsymbol{\Phi}}_i^\mathrm{H}\right)\right).\label{approximation}
\end{align}
We define the normalized trace of $\bM_{i|i-1}$ and $\bM_{i|i}$ as
\begin{align}
m_{i|i-1}&\triangleq\frac{1}{MK}\trace(\bM_{i|i-1}),\notag\\
m_{i|i}&\triangleq\frac{1}{MK}\trace(\bM_{i|i}),
\end{align}
where we denote $m_{i|i-1}$ as the prediction NMSE and $m_{i|i}$ as the minimum NMSE.

We assume that the temporal correlation coefficient is identical for all users, i.e., $\eta_k=\eta$ for all $k$. With this assumption, we can further expand $m_{i|i-1}$ and $m_{i|i}$ as
\begin{align}
m_{i|i-1}&=\frac{1}{MK}\trace(\bM_{i|i-1})\notag\\
&=\frac{1}{MK}\trace\left({\boldsymbol{\eta}}\bM_{i-1|i-1}{\boldsymbol{\eta}}^\mathrm{H}+{\boldsymbol{\zeta}}{\bR}{\boldsymbol{\zeta}}^\mathrm{H}\right)\notag\\
&=\eta^2 m_{i-1|i-1}+(1-\eta^2),\label{predictionNMSE} 
\end{align}
and
\begin{align}
&m_{i|i}\notag\\
&=\frac{1}{MK}\trace(\bM_{i|i})\notag\\
&=\frac{1}{MK}\trace\left(\left(\bI_{MK}-\bK_i\tilde{{\boldsymbol{\Phi}}}_i\right)\bM_{i|i-1}\right)\notag \\
&\stackrel{(a)}{\approx} \frac{1}{MK}\trace\Big(\Big(\bI_{MK}-\bM_{i|i-1}\tilde{{\boldsymbol{\Phi}}}_i^\mathrm{H}\Big(2\alpha\bI_{MK}\notag\\
&\qquad\qquad-\alpha^2\Big(\bC_{\tilde{\underline{\bn}}_i}+\tilde{\boldsymbol{\Phi}}_i\bM_{i|i-1}\tilde{\boldsymbol{\Phi}}_i^\mathrm{H}\Big)\Big)\tilde{{\boldsymbol{\Phi}}}_i\Big)\bM_{i|i-1}\Big)\notag \\
&\stackrel{(b)}{=} \frac{1}{MK}\trace\Big(\Big(\bI_{MK}-\bM_{i|i-1}\Big((2\alpha-\alpha^2(1-\beta))\tilde{{\boldsymbol{\Phi}}}_i^\mathrm{H}\tilde{{\boldsymbol{\Phi}}}_i\notag\\
&\qquad\qquad-\alpha^2\tilde{{\boldsymbol{\Phi}}}_i^\mathrm{H}\tilde{\boldsymbol{\Phi}}_i\bM_{i|i-1}\tilde{\boldsymbol{\Phi}}_i^\mathrm{H}\tilde{\boldsymbol{\Phi}}_i\Big)\Big)\bM_{i|i-1}\Big)\notag\\
&\stackrel{(c)}{=} \frac{1}{MK}\trace\Big(\Big(\bI_{MK}-\bM_{i|i-1}\Big((2\alpha-\alpha^2(1-\beta))\beta\notag\\
&\qquad\qquad-\alpha^2\beta\bM_{i|i-1}\beta\Big)\Big)\bM_{i|i-1}\Big)\notag\\
&\stackrel{(d)}{=}\Big(1-m_{i|i-1}\Big((2\alpha-\alpha^2(1-\beta))\beta\notag\\
&\qquad\qquad-\alpha^2\beta m_{i|i-1}\beta\Big)\Big)m_{i|i-1}\notag\\
&=(1-(2\alpha-\alpha^2(1-\beta)-\alpha^2\beta m_{i|i-1})\beta m_{i|i-1})m_{i|i-1},\label{minimumNMSE}
\end{align}
where $(a)$ is derived by the Kalman gain matrix approximation in ({\ref{approximation}}),
$(b)$ is from 
\begin{align}
\bC_{\tilde{\underline{\bn}}_i}&=\bA_i\bA_{i}^\mathrm{H}+\bC_{\bq_{i}}\notag\\
&=\left(\frac{2}{\pi}\frac{1}{K\rho+1}+\left(1-\frac{2}{\pi}\right)\right)\bI_{MK}\notag\\
&=(1-\beta)\bI_{MK},
\end{align}
$(c)$ comes from
\begin{align}
\tilde{\boldsymbol{\Phi}}_i^{\mathrm{H}}\tilde{\boldsymbol{\Phi}}_i&=\bar{\boldsymbol{\Phi}}_i^\mathrm{H}\bA_{i}^\mathrm{H}\bA_{i}\bar{\boldsymbol{\Phi}}_i\notag\\
&=\frac{2}{\pi}\frac{1}{K\rho+1}\bar{\boldsymbol{\Phi}}_i^\mathrm{H}\bar{\boldsymbol{\Phi}}_i\notag \\
&=\frac{2}{\pi}\frac{K\rho}{K\rho+1}\bI_{MK}\notag\\
&=\beta\bI_{MK},
\end{align}
and $(d)$ is derived by the fact that $\bM_{i|i-1}$ and $\bM_{i|i}$ are diagonal matrix based on the mathematical induction with $\bM_{0|0}=\bR=\bI_{MK}$. 



Now, we will show that $m_{i|i}<m_{i-1|i-1}$, i.e., the minimum NMSE decreases as the time slot index $i$ increases. 
It is enough to show that $m_{i|i-1}$ is a monotonic decreasing sequence since $m_{i|i}$ and $m_{i|i-1}$ has linear a relationship in (\ref{predictionNMSE}),
	\begin{align}
	m_{i+1|i}<m_{i|i-1} \Leftrightarrow m_{i|i}<m_{i-1|i-1}.  
	\end{align}
First, we can reformulate (\ref{predictionNMSE})  
\begin{align}
&m_{i+1|i}\notag\\
&=\eta^2 m_{i|i}+(1-\eta^2)\notag\\
         &=\eta^2 (1-(2\alpha-\alpha^2(1-\beta)-\alpha^2\beta m_{i|i-1})\beta m_{i|i-1})m_{i|i-1}\notag \\
         &\quad+(1-\eta^2).\label{m_i+1}
\end{align}
We define $f(x)$ as
\begin{align}
f(x)\triangleq\eta^2(1-(2\alpha-\alpha^2(1-\beta)-\alpha^2\beta x)\beta x)x+(1-\eta^2)\label{f(x)}.
\end{align}
Then, in Appendix B, we prove
\begin{align}
f(x)<x,~0<\gamma<x<1, \label{f_x}
\end{align}
where $\gamma$ is the root of $f(x)=x$. In (\ref{f_x}), we exploited the condition $0<\alpha<2$ that is proved in Appendix C. 
Thus, we conclude 
\begin{align}
m_{i+1|i}=f(m_{i|i-1})<m_{i|i-1},
\end{align}
which is equivalent to $m_{i|i}<m_{i-1|i-1}$. Furthermore, we prove 
\begin{align}
\lim_{i\rightarrow \infty} m_{i|i-1}=\gamma, \label{limit}
\end{align}in Appendix D. Therefore, the prediction NMSE $m_{i|i-1}$ decreases as the time slot index $i$ increases and converges to $\gamma$. Also, we can easily check that $m_{1|1}=1-(2\alpha-\alpha^2)\beta=1-\beta=\mathrm{NMSE}_{\mathrm{BLM}}$ with $m_{1|0}=1$ and $\alpha=1$. After many time instances, we will have
\begin{align} 
\mathrm{NMSE}_{\mathrm{BLM}}=m_{1|1}\gg m_{i|i},
\end{align}
and the TPE-based estimator would outperform the BLMMSE estimator.

So far, we assume that $\bR=\bI_{MK}$, i.e., spatially uncorrelated channels, to derive the NMSE of the TPE estimator. Even for spatially correlated channels, the numerical results in Section {\ref{sec:numericalresult}} show that the TPE-based estimator outperforms the BLMMSE estimator.

\section{Uplink Data Transmission}\label{sec:datatransmission}
{In this section, we derive the achievable sum-rate of massive MIMO with one-bit ADCs following similar steps as in \cite{Li17} for the sake of completeness.} The $K$ users transmit data symbols to the BS. Based on the Bussgang decomposition, the quantized signal in the $i$-th time slot can be represented as
\begin{align}
\br_{d,i}\notag&=\cQ(\sqrt{\rho_{d,i}}\bH_{i}\bs_{i}+\bn_{d,i})\\
&=\sqrt{\rho_{d,i}}\bA_{d,i}\bH_i\bs_i+\bA_{d,i}\bn_{d,i}+\bq_{d,i},\label{receivedata}
\end{align}
where $\bs_i$ is the transmit signal satisfying 
${\mathbb{E}}\{|s_{i,k}|^2\}=1$, and the subscript $d$ denotes the data transmission. 
The linear operator in (\ref{receivedata}) can be approximated as
\begin{align}
\bA_{d,i}&=\sqrt{\frac{2}{\pi}}\diag(\bC_{\by_{d,i}})^{-\frac{1}{2}}\notag\\
&=\sqrt{\frac{2}{\pi}}\diag(\rho_{d,i}\bH_i\bH_i^\mathrm{H}+\bI_M)^{-\frac{1}{2}}\notag\\
&\stackrel{(a)}{\approx}{\sqrt{\frac{2}{\pi}}\sqrt{\frac{1}{K\rho_{d,i}+1}}\bI_{M}}.\label{linearoperatordata}
\end{align}
In (\ref{linearoperatordata}), $(a)$ is from the channel hardening effect in massive MIMO systems as in \cite{Li17}. After applying the receive combiner for the quantized signal, we have 
\begin{align}
\hat\bs_i\notag&=\bW_i^\mathrm{T}\br_{d,i}\\
&=\sqrt{\rho_{d,i}}\bW_i^\mathrm{T}\bA_{d,i}(\hat{\bH}_i\bs_i+\boldsymbol{\cE}_i\bs_i)+\bW_i^\mathrm{T}\bA_{d,i}\bn_{d,i}+\bW_i^\mathrm{T}\bq_{d,i},
\end{align}
where $\bW_i$ is the receive combining matrix, $\hat{\bH}_i =\text{unvec}({\hat{\underline{\bh}}_i})$ is the unvectorized channel estimation matrix, and $\boldsymbol{\cE}_i=\bH_i-\hat{\bH}_i$ is the estimation error matrix.
The $k$-th element of $\hat{\bs}_i$ can be represented as
\begin{align}
&\hat{s}_{i,k}\notag\\
&=\sqrt{\rho_{d,i}}\bw_{i,k}^\mathrm{T}\bA_{d,i}\hat{\bh}_{i,k}s_{i,k}+\sqrt{\rho_{d,i}}\bw_{i,k}^\mathrm{T}\sum_{j\neq k}^K\bA_{d,i}\hat{\bh}_{i,j} s_{i,j}\notag\\
&+\sqrt{\rho_{d,i}}\bw_{i,k}^\mathrm{T}\sum_{j=1}^K\bA_{d,i}\boldsymbol{\epsilon}_{i,j} s_{i,j}+\bw_{i,k}^\mathrm{T}\bA_{d,i}\bn_{d,i}+\bw_{i,k}^\mathrm{T}\bq_{d,i},
\end{align}
where $\bw_{i,k},\hat{\bh}_{i,k}$ and $\boldsymbol{\epsilon}_{i,k}$ represent the {$k$-th} columns of $\bW_i,\hat{\bH}_i$, and $\boldsymbol{\cE}_i$, respectively.

We can obtain a lower bound on the achievable rate of the $k$-th user by treating the uncorrelated inter-user interference (IUI) and the quantization noise (QN) $\bq_{d,i}$ as a Gaussian noise \cite{Diggavi01}, and assuming the Gaussian channel input as in \cite{Li17},
\begin{align}
R_{i,k}=\mathbb{E} \left\{ \log_2\left( 1+\frac{\mathrm{S}_{i,k}}{\mathrm{IUI}_{i,k}+\mathrm{QN}_{i,k}} \right) \right\},\label{Rate}
\end{align}
where 
\begin{align}
\mathrm{S}_{i,k}&=\rho_{d,i}|\bw_{i,k}^\mathrm{T}\bA_{d,i}\hat{\bh}_{i,k}|^2,\notag \\
\mathrm{IUI}_{i,k}&=\rho_{d,i}\sum_{j \neq k}^K|\bw_{i,k}^\mathrm{T}\bA_{d,i}\hat{\bh}_{i,j}|^2,\notag \\
\mathrm{QN}_{i,k}&=\rho_{d,i}\sum_{j=1}^K|\bw_{i,k}^\mathrm{T}\bA_{d,i}\boldsymbol{\epsilon}_{i,j}|^2+\norm{\bw_{i,k}^\mathrm{T}\bA_{d,i}}^2\notag\\
&\quad+\bw_{i,k}^\mathrm{T}\bC_{\bq_{d,i}}\bw_{i,k}^*.
\end{align}
The auto-covariance matrix of $\bq_{d,i}$ is given by
\begin{align}
{\bC_{\bq_{d,i}}}\notag&=\bC_{\br_{d,i}}-\bA_{d,i}\bC_{\by_{d,i}}\bA_{d,i}^\mathrm{H}\\
\notag&=\frac{2}{\pi}(\text{arcsin}(\bX_{d,i})+j \text{arcsin}(\bY_{d,i}))-\frac{2}{\pi}(\bX_{d,i}+j\bY_{d,i})\notag \\
&\stackrel{(a)}{\approx}(1-2/\pi)\bI_{M},\label{covariancematrixq}
\end{align}
where we define
\begin{align}
\bX_{d,i}&=\Sigma_{\by_{d,i}}^{-1/2}\Re\{\bC_{\by_{d,i}}\}\Sigma_{\by_{d,i}}^{-1/2},\notag\\
\bY_{d,i}&=\Sigma_{\by_{d,i}}^{-1/2}\Im\{\bC_{\by_{d,i}}\}\Sigma_{\by_{d,i}}^{-1/2}.
\end{align}
In (\ref{covariancematrixq}), $\bC_{\br_{d,i}}$ can be obtained by the arcsin law in (\ref{arcsinlaw}), and $(a)$ comes from the approximation of the low SNR as in \cite{Li17}. This approximation holds even in correlated channels, which is different from (\ref{covariancematrixqi}) that is based on the assumption $\bR=\bI_{MK}$. 
We define the achievable sum-rate as
\begin{align}
R_i=\sum_{k=1}^K{R_{i,k}}.
\end{align}
To reduce the interference, we adopt the zero-forcing (ZF) combiner, 
\begin{align}
{\bW_{i,\mathrm{ZF}}^\mathrm{T}=(\hat{\bH}_i^\mathrm{H}\hat{\bH}_i)^{-1}\hat{\bH}_i^\mathrm{H},}
\end{align}
for numerical studies.

\section{Results and Discussion}\label{sec:numericalresult}
In this section, we verify the proposed channel estimator by Monte-Carlo simulation. We define the NMSE as the performance metric,
\begin{align}
{\mathrm{NMSE}}=
\frac{1}{MK}{\mathbb{E}}\left\{ \left\|\hat{\underline{\bh}}-\underline{\bh}\right\|_2^2\right\},
\end{align}
where $\underline{\hat{\bh}}$ is the channel estimate and $\underline{\bh}$ is the true channel.
We adopt the pilot matrix $\boldsymbol{\Phi}$ by the discrete Fourier transform (DFT) matrix, which satisfies the assumptions in Section \ref{single}, and select $K$ columns of $\tau\times\tau$ DFT matrix with $\tau \geq K$ to obtain the pilot sequences.
We adopt the Jakes' model for the temporal correlation, which is given as $\eta_k=J_0(2\pi f_{D,k}t)$ where $J_0(\cdot)$ denotes the $0$-th order Bessel function, $f_{D,k}$ is the Doppler frequency, and $t$ is the channel instantiation interval. For simulations, we set $f_{D,k}=v_kf_c/c$ with the user speed $v_k$, the carrier frequency $f_c=2.5~\text{GHz}$, and the speed of light $c=3\times 10^8 ~\mathrm{m}\cdot \mathrm{s}^{-1}$. We also set $t=5~\mathrm{ms}$ \cite{Choi12}.
We denote ${\mathrm{NMSE}}(\bh_i)$ as the NMSE of KFB estimator at the $i$-th time slot and   ${\mathrm{NMSE}}(\bM)=\frac{1}{MK}\trace(\bM_{i|i})$ as the theoretical NMSE of Kalman filtering with the Gaussian noise, not the quantization noise. Therefore, ${\mathrm{NMSE}}(\bM)$ gives the performance limit of Kalman filtering with the Gaussian noise. We depict the ``BLMMSE'' as the NMSE performance of the single-shot channel estimator discussed in Section \ref{single}.

\begin{figure}[t]
	\centering
	\includegraphics[width=9 cm]{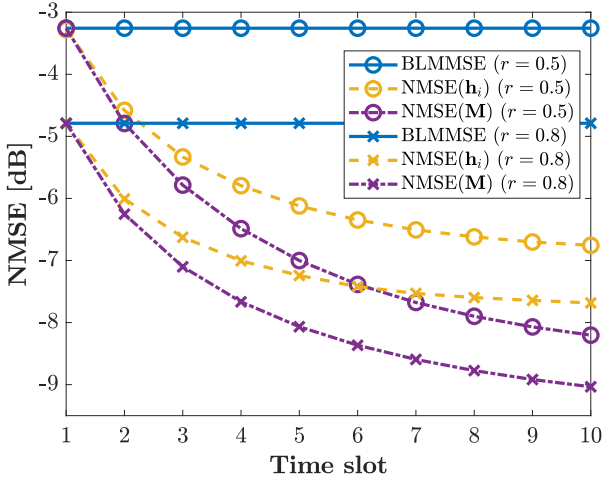}
	\caption{The NMSEs of BLMMSE estimator, KFB estimator, and theoretical limit of Kalman filtering according to time slot $i$ for different spatial correlation coefficient $r$ when {$M=128$}, $K=8$, $\tau=8$, $\eta_k=0.988$, and SNR = $-5$ dB.}\label{fig:NMSE1}
\end{figure}

\begin{figure}[t]
	\centering
	\includegraphics[width=9 cm]{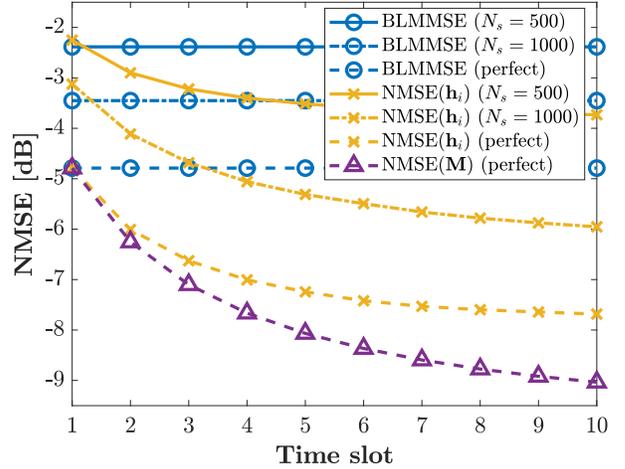}
	\caption{The NMSEs of BLMMSE estimator, KFB estimator, and theoretical limit of Kalman filtering according to time slot $i$ with and without the perfect spatial correlation knowledge when $M=128$, $K=8$, $\tau=8$, $\eta_k=0.988$, $r=0.8$, and SNR = $-5$ dB.}\label{fig:NMSE11}
\end{figure}

\begin{figure}[t]
	\centering
	\includegraphics[width=9 cm]{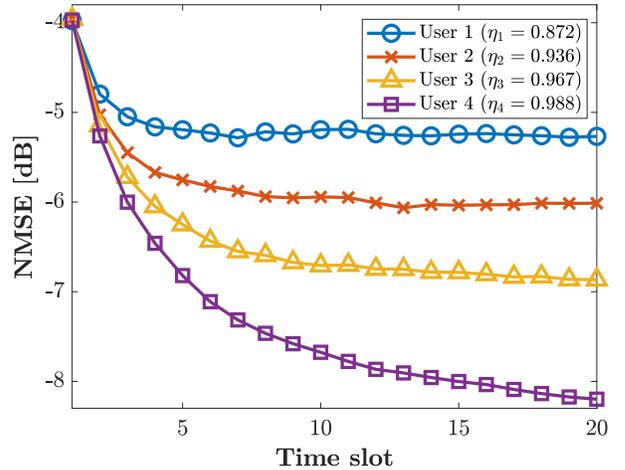}
	\caption{The NMSEs of KFB estimator according to time slot $i$ for different temporal fading users when $M=128$, $K=4$, $\tau=4$, $r=0.8$, and SNR = $-5$ dB.}\label{fig:NMSE2}
\end{figure}



In Fig. \ref{fig:NMSE1}, we compare the NMSEs of BLMMSE estimator and KFB estimator with the time slot $i$ for $r=0.5$ or $r=0.8$ with SNR = $-5$ dB. We assume the BS antennas {$M=128$}, the users {$K=8$}, and the symbols {$\tau=8$}. We set the temporal correlation coefficient $\eta_k=0.988$, which corresponds to $v=3~\mathrm{km/h}$. {As the time slot} increases, the proposed KFB estimator outperforms the BLMMSE estimator. By comparing ${\mathrm{NMSE}}(\bh_i)$ and ${\mathrm{NMSE}}(\bM)$, the loss from using one-bit ADCs is around $1.5$ dB. As the amount of spatial correlation increases from $0.5$ to $0.8$, all estimators perform better since it becomes easier to estimate channels as the channels become more correlated in space \cite{Kotecha04,Bjornson10,Choi142}.

In Fig. \ref{fig:NMSE11}, we compare the NMSEs of the channel estimators with and without the perfect spatial correlation knowledge. All the parameters are the same as in Fig. \ref{fig:NMSE1} with $r=0.8$. Without the spatial correlation knowledge, we use $N_s$ samples to estimate the spatial channel correlation by the LS estimates, then we estimate the channel. When we use $N_s=500, 1000$, the performance loss is about $4,2$ dB compare to the case of perfect correlation knowledge. Although the performance degradation due to the imperfect knowledge of spatial correlation is non-negligible, the loss is inevitable for the channel estimators,	including the BLMMSE estimator, that exploit the spatial correlation. The KFB estimator outperforms the BLMMSE estimator even with the sample correlation matrix, and as time slot increases, the KFB estimator using the sample correlation matrix achieves lower NMSE than the BLMMSE estimator using the true spatial correlation matrix.

Fig. \ref{fig:NMSE2} depicts the NMSEs of the KFB estimator when each user experiences different temporal fading. We set $r=0.8$ and the temporal correlation coefficient of user 1 to 4 as $\eta_k=0.872, 0.936, 0.967$, and $0.988$, which correspond to $v_k=10~\mathrm{km/h}, 7~\mathrm{km/h}, 5~\mathrm{km/h}$, and $3~\mathrm{km/h}$. All other settings are the same as in Fig. \ref{fig:NMSE1}. As expected, the users with high temporal correlations benefit more from the KFB estimator. Even the user with the moderate velocity of $10~\mathrm{km/h}$ also has the gain more than $1$ dB.

\begin{figure}[t]
	\centering
	\includegraphics[width=9 cm]{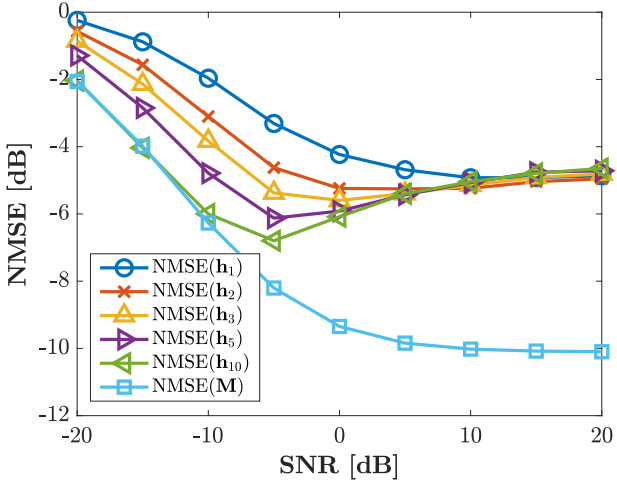}
	\caption{The NMSEs of KFB estimator according to SNR with different time slots when $M=128$, $K=8$, $\tau=8$, $\eta_k=0.988$, and $r=0.5$.}\label{fig:NMSE3}
\end{figure}

\begin{figure}[t]
	\centering
	\includegraphics[width=9 cm]{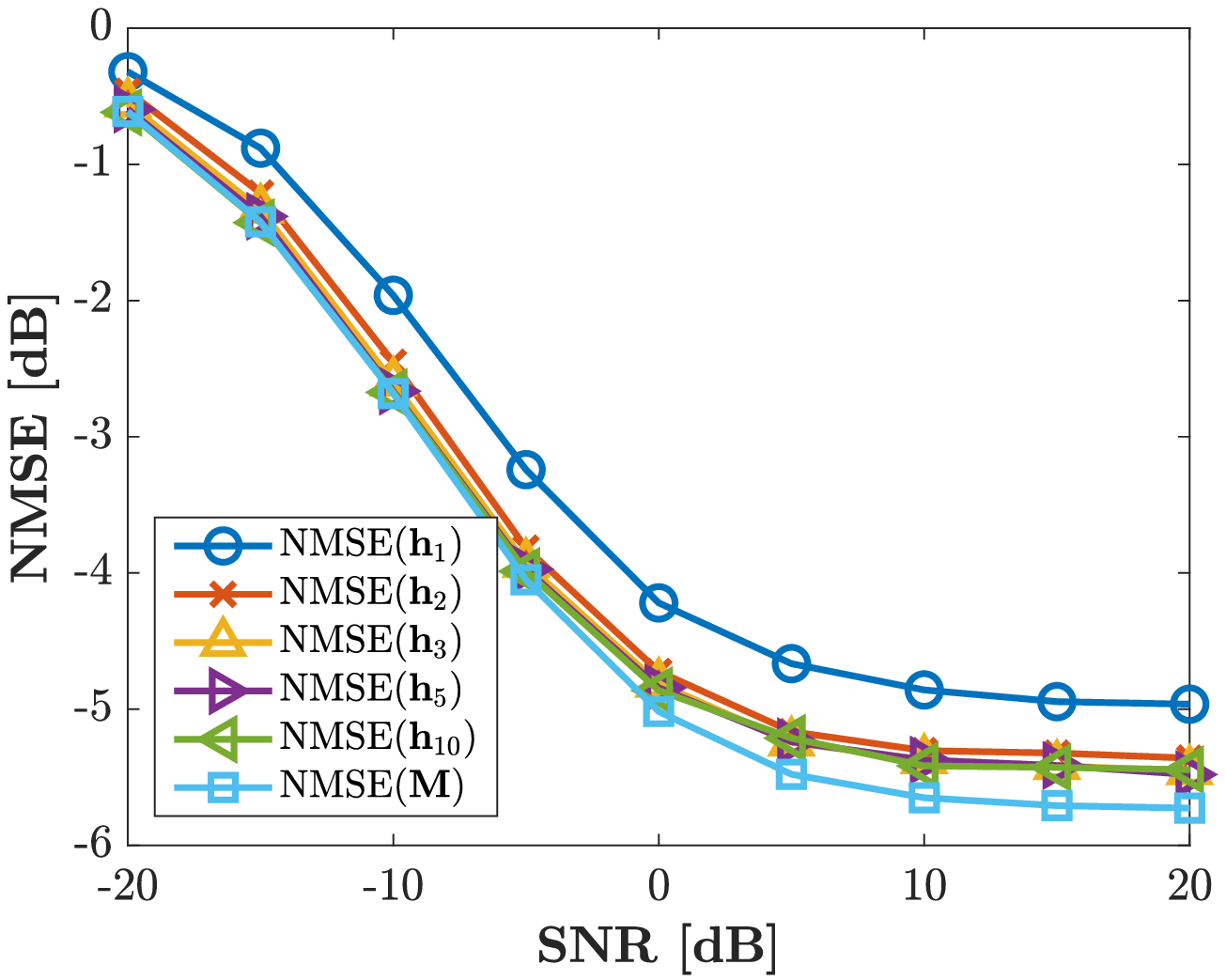}
	\caption{The NMSEs of KFB estimator according to SNR with different time slots when {$M=128$}, $K=8$, $\tau=8$, $\eta_k=0.724$, and $r=0.5$.}\label{fig:NMSE32}
\end{figure}

\begin{figure}[t]
	\centering
	\includegraphics[width=9 cm]{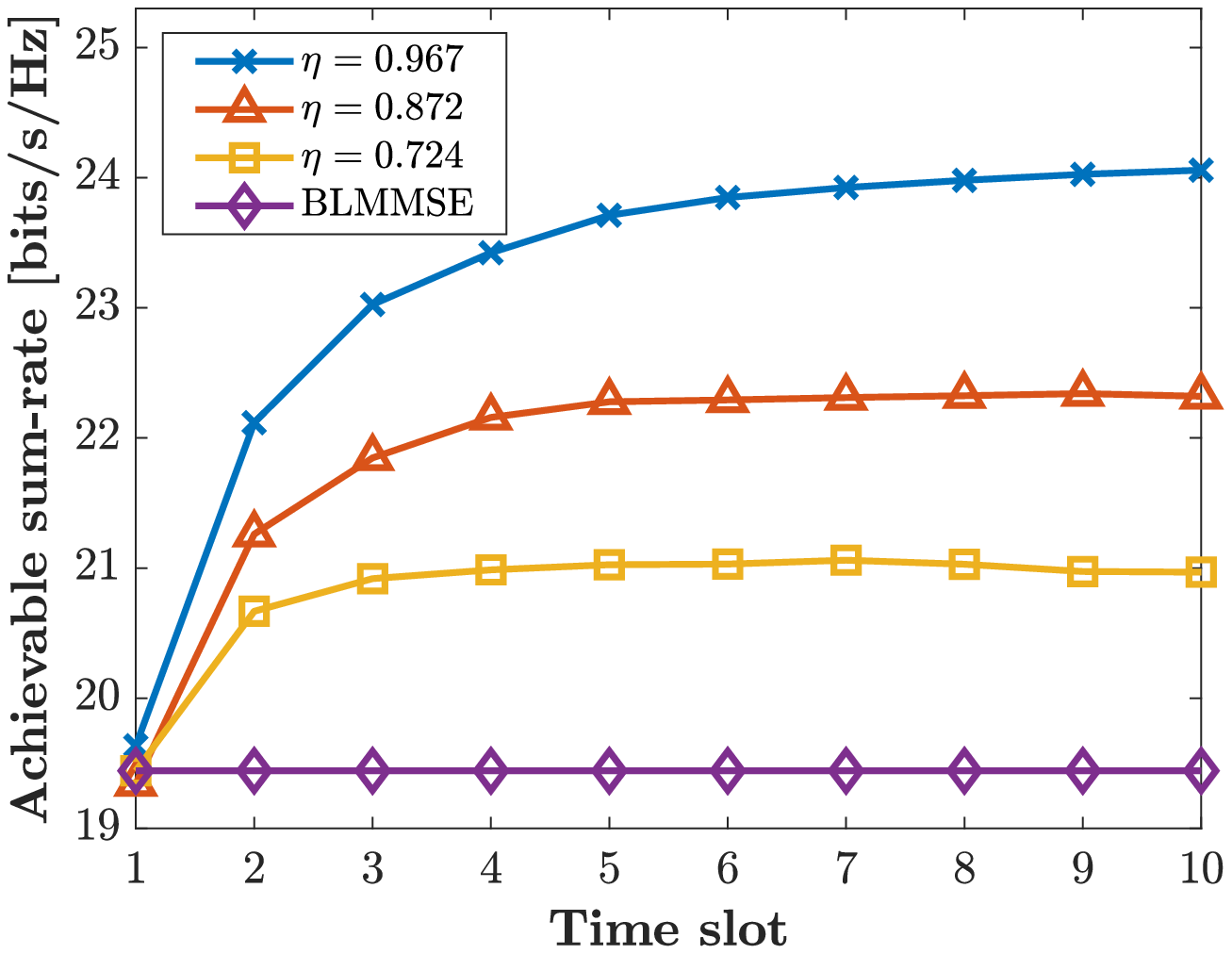}
	\caption{The achievable sum-rates of BLMMSE estimator and KFB estimator according to time slot $i$ with different temporal correlations when $M=128$, $K=8$, $\tau=8$,  $r=0.8$, and SNR = $0$ dB.}\label{fig:Sumrate_Time_SNR0}
\end{figure}

\begin{figure}[t]
	\centering
	\includegraphics[width=9 cm]{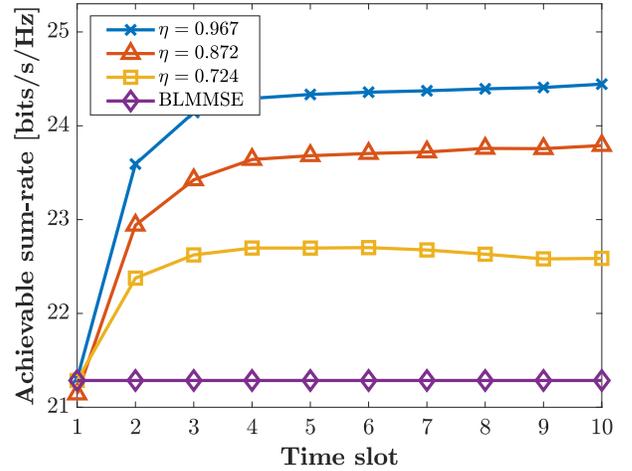}
	\caption{The achievable sum-rates of BLMMSE estimator and KFB estimator according to time slot $i$ with different temporal correlations when $M=128$, $K=8$, $\tau=8$,  $r=0.8$, and SNR = $10$ dB.}\label{fig:Sumrate_Time_SNR10}
\end{figure}

In Figs. \ref{fig:NMSE3} and \ref{fig:NMSE32}, we compare the NMSEs of the KFB estimator according to SNR with different time slots when $M=128$, $K=8$, $\tau=8$, $\eta_k=0.988, 0.724$ (correspond to $v_k=5,15$ km/h), and $r=0.5$. When the temporal correlation is high, the NMSEs of the KFB estimator decreased as the SNR increased in low SNR regime. In the low SNR regime, ${\mathrm{NMSE}}(\bh_i)$ is almost the same the theoretical NMSE of ${\mathrm{NMSE}}(\bM)$ after 10 successive estimations. In the high SNR, however, the NMSE of KFB estimator suffers from the saturation effect, which is referred as the stochastic resonance due to one-bit quantization noise \cite{Dither14}. In the proposed KFB estimator, the loss also comes from the Gaussian model mismatch in the one-bit quantization as explained in \textbf{Remark 1} in Section \ref{Successive}.
When the temporal correlation is low, the NMSEs of the KFB estimator decreased as the SNR increased in all SNR regime. This is because the channel estimation error comes mostly from the large temporal channel variation, not from the one-bit quantization. 

In Figs. \ref{fig:Sumrate_Time_SNR0} and \ref{fig:Sumrate_Time_SNR10}, we compare the achievable sum-rates of the BLMMSE estimator and KFB estimator according to the time slot when $M=128$, $K=8$, $\tau=8$, $r=0.8$, and SNR = $0$ and $10$ dB. We assume all users experience the same $\eta$. In both scenarios, the achievable sum-rate of the KFB estimator outperforms the BLMMSE estimator as the time slot increases.

\begin{figure}[t]
	\includegraphics[width=9.1 cm]{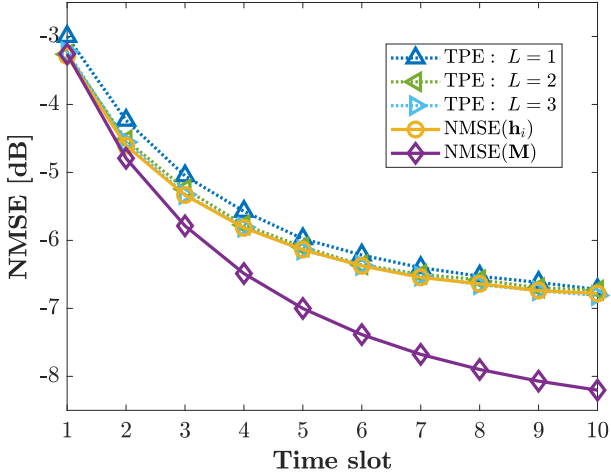}
	\caption{The NMSEs of KFB estimator and TPE-based estimator according to time slot $i$ when $M=128$, $K=8$, $\tau=8$, $\eta_k=0.988$, $r=0.5$, $\alpha=0.5$, and SNR = $-5$ dB.}\label{fig:NMSE_TIME_TPE}
\end{figure}

\begin{figure}[t]
	\includegraphics[width=9.1 cm]{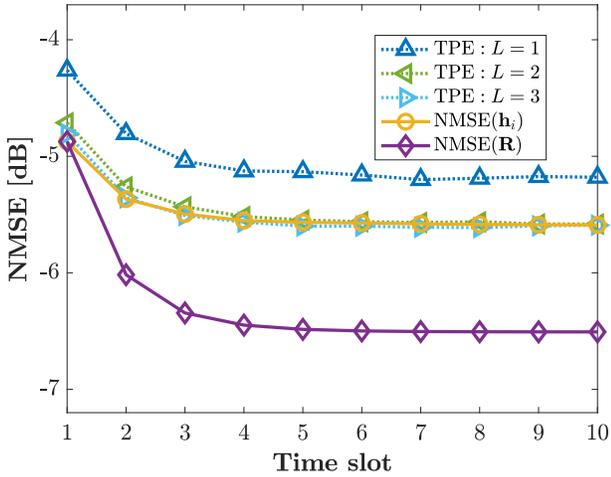}
	\caption{The NMSEs of KFB estimator and TPE-based estimator according to time slot $i$ when $M=128$, $K=8$, $\tau=8$, $\eta_k=0.872$, $r=0.5$, $\alpha=0.5$, and SNR = $10$ dB.}\label{fig:NMSE_TIME_TPE2}
\end{figure}

In Figs. \ref{fig:NMSE_TIME_TPE} and \ref{fig:NMSE_TIME_TPE2}, we compare the NMSEs of the KFB estimator and the low-complexity TPE-based estimator with the time slot. We set $M=128$, $K=8$, $\tau=8$, $r=0.5$, $\eta_k=0.988, 0.872$, and SNR = $-5, 10$ dB. We numerically optimize $\alpha=0.5$ for the TPE-based estimator. In the high temporal correlation and low SNR case (Fig. \ref{fig:NMSE_TIME_TPE}), the NMSE gap between the KFB and TPE-based estimators is negligible and already quite small even with $L=1$.
In the low temporal correlation and high SNR case (Fig. \ref{fig:NMSE_TIME_TPE2}), the performance is degraded but the gap becomes small with $L=2$. Therefore, in practice, the low-complexity TPE-based estimator can be used with negligible performance loss.

\section{Conclusion}\label{sec:conclusions}
In this paper, we proposed the Kalman filter-based (KFB) channel estimators that exploit both the spatial and temporal correlations of channels for massive MIMO systems using one-bit ADCs. We adopted the Bussgang decomposition to linearize the non-linear effect from one-bit quantization. Based on the linearized model and assuming the effective noise as Gaussian, we exploited the Kalman filter to estimate the channel successively. The proposed KFB estimator has a remarkable gain compared to the previous estimator in \cite{Li17}, which does not exploit any temporal correlation in channels. To resolve the complexity issue of the KFB estimator due to the large-scale matrix inversion, we also implemented the truncated polynomial expansion (TPE)-based estimator. We analytically derived the minimum NMSE based on the first-order TPE approximation, and the numerical results showed that the low-complexity TPE-based estimator gives nearly the same accuracy as the KFB estimator even with lower approximation orders.

\begin{appendices}
\section{Proof of (\ref{A})}
We first expand $\bar{\boldsymbol{\Phi}}{\bR}\bar{\boldsymbol{\Phi}}^\mathrm{H}$ as,
\begin{align}
&\bar{\boldsymbol{\Phi}}{\bR}\bar{\boldsymbol{\Phi}}^\mathrm{H}\notag\\
&=(\boldsymbol{\Phi}\otimes \sqrt{\rho}\bI_M){\bR}(\boldsymbol{\Phi}\otimes\sqrt{\rho}\bI_M)^\mathrm{H}\notag\\
&=\rho\left(\begin{bmatrix} \phi_{1,1} &\cdots &\phi_{1,K} \\
\vdots     &\ddots & \vdots \\
\phi_{\tau,1} & \cdots &\phi_{\tau,K}
\end{bmatrix} \otimes \bI_M\right) {\bR}(\boldsymbol{\Phi}\otimes\bI_M)^\mathrm{H}\notag\\
&=\rho\begin{bmatrix} \phi_{1,1}\bI_M &\cdots &\phi_{1,K} \bI_M\\
\vdots     &\ddots & \vdots \\
\phi_{\tau,1}\bI_M & \cdots &\phi_{\tau,K}\bI_M
\end{bmatrix}{\bR}(\boldsymbol{\Phi}\otimes\bI_M)^\mathrm{H}\notag\\
&\stackrel{(a)}{=}\rho\begin{bmatrix} \phi_{1,1}\bR_1 &\cdots &\phi_{1,K} \bR_K\\
\vdots     &\ddots & \vdots \\
\phi_{\tau,1}\bR_1 & \cdots &\phi_{\tau,K}\bR_K
\end{bmatrix}(\boldsymbol{\Phi}\otimes\bI_M)^\mathrm{H}\notag\\
&=\rho\begin{bmatrix} \phi_{1,1}\bR_1 &\cdots &\phi_{1,K} \bR_K\\
\vdots     &\ddots & \vdots \\
\phi_{\tau,1}\bR_1 & \cdots &\phi_{\tau,K}\bR_K
\end{bmatrix}\begin{bmatrix} \phi_{1,1}^*\bI_M &\cdots &\phi_{\tau,1}^* \bI_M\\
\vdots     &\ddots & \vdots \\
\phi_{1,K}^*\bI_M & \cdots &\phi_{\tau,K}^*\bI_M
\end{bmatrix}\notag\\
&\stackrel{(b)}{=}\rho\begin{bmatrix} \sum_{k=1}^K\bR_k &\cdots &~\\
\vdots     &\ddots & \vdots \\
~ & \cdots &\sum_{k=1}^K\bR_k
\end{bmatrix},
\end{align}
where $(a)$ comes from the independent spatial correlation matrix $\bR$ in (\ref{R_k}), and $(b)$ follows by assuming that the pilot sequences are column-wise orthogonal with the same magnitude for all elements. Since the diagonal term of $\bR_k$ is 1 for all $k$, we have
\begin{align}
\diag\left(\bar{\boldsymbol{\Phi}}{\bR}\bar{\boldsymbol{\Phi}}^\mathrm{H}\right)=K\rho\bI_{M\tau},
\end{align}
which finishes the proof.

\section{Proof of (\ref{f_x})}
First, we define $g(x)$ as
	\begin{align}
	g(x)&\triangleq f(x)-x\notag\\
	    &=\eta^2(\alpha^2\beta^2x^3-(2\alpha-\alpha^2+\alpha^2\beta)\beta x^2 +x) \notag\\
	    &\quad+1-\eta^2-x.        
	\end{align}
Then, we have
	\begin{align}
	g(-\infty)&<0,\notag\\
	g(0)&=1-\eta^2>0,\notag\\
	g(1)&=\eta^2\beta(-2\alpha+\alpha^2)<0,\notag\\
	g(\infty)&>0,
	\end{align}
	where the inequality of $g(1)$ is due to $0<\alpha<2$. The roots of $g(x)$ are $\gamma_{0^{-}}\in(-\infty,0)$, $\gamma\in(0,1)$, $\gamma_{1^+}\in(1,\infty)$ since $g(x)$ is the third-order polynomial, and $g(-\infty)g(0)<0$, $g(0)g(1)<0$, $g(1)g(\infty)<0$ based on the intermediate value theorem. Therefore, $\gamma$ is the unique solution of $g(x)=0$ on $x\in(0,1)$. 

Now, we will show $g(x)<0$ for $0<\gamma<x<1$. The derivative of $g(x)$ is given by
\begin{align}
g'(x)&=f'(x)-1\notag\\
     &=\eta^2\Big(3\alpha^2\beta^2x^2-2(2\alpha-\alpha^2+\alpha^2\beta)\beta x+1\Big)-1.
\end{align}
Then, we have
\begin{align}
g'(0)&=\eta^2-1<0, \notag\\
g'(1)&=\eta^2(\alpha^2\beta^2-4\alpha\beta+2\alpha^2\beta+1)-1\notag\\
     &=\eta^2\alpha\beta(\alpha(\beta+2)-4)+\eta^2-1\notag \\
     &<\eta^2\alpha\beta(\alpha(\beta+2)-4)\notag \\
     &<\eta^2\alpha\beta(2\alpha-4)\notag \\
     &<0,
\end{align}
since $0<\alpha<2$. This result implies $g'(x)<0$ for $0<x<1$ because $g'(x)$ is the second-order polynomial with the positive leading coefficient $3\eta^2\alpha^2\beta^2$.
Since $g'(x)<0$ for $0<x<1$ and $g(\gamma)=0$, then $g(x)<g(\gamma)=0$ for $0<\gamma<x<1$, which finishes the proof.

\section{Proof of $0<\alpha<2$}
We first reformulate $\max_n\lambda_n\Big(\bC_{\tilde{\underline{\bn}}_i}+\tilde{\boldsymbol{\Phi}}_i\bM_{i|i-1}\tilde{\boldsymbol{\Phi}}_i^\mathrm{H}\Big)$ as,
	\begin{align}
	\max_n\lambda_n\Big(\bC_{\tilde{\underline{\bn}}_i}+\tilde{\boldsymbol{\Phi}}_i\bM_{i|i-1}\tilde{\boldsymbol{\Phi}}_i^\mathrm{H}\Big)&{\stackrel{(a)}= 1-\beta+\beta m_{i|i-1}}\notag\\
	&=1-\beta(1-m_{i|i-1})\notag\\
	&{\stackrel{(b)}\leq1,} \label{convergence}
	\end{align}
where $(a)$ comes from the fact that $\bM_{i|i-1}$ is a diagonal matrix, and $(b)$ is from $0<\gamma\leq m_{i|i-1}\leq1$.
By plugging (\ref{convergence}) into the bound on $\alpha$,
	\begin{align}
	0<\alpha<\frac{2}{\max_n \lambda_n\Big(\bC_{\tilde{\underline{\bn}}_i}+\tilde{\boldsymbol{\Phi}}_i\bM_{i|i-1}\tilde{\boldsymbol{\Phi}}_i^\mathrm{H}\Big)},
	\end{align}
	we have the tightened bound
	\begin{align}
	0<\alpha<2\leq\frac{2}{\max_n \lambda_n\Big(\bC_{\tilde{\underline{\bn}}_i}+\tilde{\boldsymbol{\Phi}}_i\bM_{i|i-1}\tilde{\boldsymbol{\Phi}}_i^\mathrm{H}\Big)}.
	\end{align}

\section{Proof of (\ref{limit})}
Based on the mathematical induction, we assume 
\begin{align}
m_{1|0}=1> \gamma,\notag\\
m_{i|i-1}> \gamma.
\end{align}
First, we proof that $f(x)$ is the increasing function on $x\in(\gamma,1)$.
The derivative of $f(x)$ is 
\begin{align}
f'(x)=\eta^2\Big(3\alpha^2\beta^2x^2-2(2\alpha-\alpha^2+\alpha^2\beta)\beta x+1\Big).
\end{align}
Then, we have
\begin{align}
f'(0)&=\eta^2>0,\notag\\
f'(1)&=\eta^2(\alpha^2\beta^2-4\alpha\beta+2\alpha^2\beta+1)\notag\\
     &=\eta^2((\beta^2+2\beta)\alpha^2-4\beta\alpha+1)\notag\\
     &=\eta^2\left((\beta^2+2\beta)\alpha-\frac{2\beta}{\beta^2+2\beta}\right)^2+\frac{\beta(2-3\beta)}{\beta^2+2\beta}\notag \\
     &>0,
\end{align}
which comes from $0<\beta<\frac{2}{\pi}<\frac{2}{3}$.
Thus, $f(x)$ is the increasing function on $x\in(\gamma,1)$. Finally,
we get $f(x)>f(\gamma)=\gamma$, which implies $m_{i+1|i}=f(m_{i|i-1})>f(\gamma)=\gamma$. Thus, $m_{i|i-1}>\gamma$ for all $i>0$ due to the mathematical induction. Since $m_{i|i-1}$ is the monotonic decreasing and bounded sequence, $m_{i|i-1}$ converges by the monotone convergence theorem \cite{Bibby74}.

Thus, we can define $L_m=\lim_{i\rightarrow \infty}m_{i|i-1}$, 
\begin{align}
L_m&=\lim_{i\rightarrow \infty} m_{i+1|i}\notag\\
   &=\lim_{i\rightarrow \infty} f(m_{i|i-1}) \notag \\
   &=f(L_m),
\end{align}
which implies $L_m$ is also a root of $f(x)=x$. 
Since $m_{i|i-1}$ converges $L_m$, and $\gamma$ is the unique solution of $f(x)=x$ on $x\in(0,1)$, $L_m=\lim_{i\rightarrow \infty} m_{i|i-1}=\gamma$, which finishes the proof.

\end{appendices}




\bibliographystyle{IEEEtran}
\bibliography{Eurasip}

\begin{thebibliography}{10}
\providecommand{\url}[1]{#1}
\csname url@samestyle\endcsname
\providecommand{\newblock}{\relax}
\providecommand{\bibinfo}[2]{#2}
\providecommand{\BIBentrySTDinterwordspacing}{\spaceskip=0pt\relax}
\providecommand{\BIBentryALTinterwordstretchfactor}{4}
\providecommand{\BIBentryALTinterwordspacing}{\spaceskip=\fontdimen2\font plus
\BIBentryALTinterwordstretchfactor\fontdimen3\font minus
  \fontdimen4\font\relax}
\providecommand{\BIBforeignlanguage}[2]{{%
\expandafter\ifx\csname l@#1\endcsname\relax
\typeout{** WARNING: IEEEtran.bst: No hyphenation pattern has been}%
\typeout{** loaded for the language `#1'. Using the pattern for}%
\typeout{** the default language instead.}%
\else
\language=\csname l@#1\endcsname
\fi
#2}}
\providecommand{\BIBdecl}{\relax}
\BIBdecl

\bibitem{2018Globecom}
H.~{Kim} and J.~{Choi}, ``Channel estimation for one-bit massive {MIMO} systems
  exploiting spatio-temporal correlations,'' in \emph{2018 IEEE Global
  Communications Conference (GLOBECOM)}, Dec. 2018, pp. 1--6.

\bibitem{Marzetta06}
T.~L. Marzetta, ``Noncooperative {c}ellular {w}ireless with {u}nlimited
  {n}umbers of {b}ase {s}tation {a}ntennas,'' \emph{IEEE Transactions on
  Wireless Communications}, vol.~9, no.~11, pp. 3590--3600, Nov. 2010.

\bibitem{Rusek13}
F.~Rusek, D.~Persson, B.~K. Lau, E.~G. Larsson, T.~L. Marzetta, O.~Edfors, and
  F.~Tufvesson, ``{S}caling {u}p {MIMO}: {o}pportunities and {c}hallenges with
  {v}ery {l}arge {a}rrays,'' \emph{IEEE Signal Processing Magazine}, vol.~30,
  no.~1, pp. 40--60, Jan. 2013.

\bibitem{Hoydis13}
J.~Hoydis, S.~ten Brink, and M.~Debbah, ``Massive {MIMO} in the {UL/DL} of
  {c}ellular {n}etworks: {h}ow {m}any {a}ntennas {d}o {w}e {n}eed?'' \emph{IEEE
  Journal on Selected Areas in Communications}, vol.~31, no.~2, pp. 160--171,
  Feb. 2013.

\bibitem{Larsson14}
E.~G. Larsson, O.~Edfors, F.~Tufvesson, and T.~L. Marzetta, ``{M}assive {MIMO}
  for next generation wireless systems,'' \emph{IEEE Communications Magazine},
  vol.~52, no.~2, pp. 186--195, Feb. 2014.

\bibitem{Walden99}
R.~H. Walden, ``Analog-to-digital converter survey and analysis,'' \emph{IEEE
  Journal on Selected Areas in Communications}, vol.~17, no.~4, pp. 539--550,
  Apr. 1999.

\bibitem{Wang15}
{S. Wang and Y. Li and J. Wang}, ``Multiuser {d}etection in {m}assive {s}patial
  {m}odulation {MIMO} {w}ith {l}ow-{r}esolution {ADC}s,'' \emph{IEEE
  Transactions on Wireless Communications}, vol.~14, no.~4, pp. 2156--2168,
  Apr. 2015.

\bibitem{Wang14}
S.~Wang, Y.~Li, and J.~Wang, ``{C}onvex optimization based multiuser detection
  for uplink large-scale {MIMO} under low-resolution quantization,'' in
  \emph{2014 IEEE International Conference on Communications (ICC)}, Jun. 2014,
  pp. 4789--4794.

\bibitem{Zhang16}
T.~Zhang, C.~Wen, S.~Jin, and T.~Jiang, ``{M}ixed-{ADC} massive {MIMO}
  detectors: performance analysis and design optimization,'' \emph{IEEE
  Transactions on Wireless Communications}, vol.~15, no.~11, pp. 7738--7752,
  Nov. 2016.

\bibitem{Jeon17}
Y.~Jeon, S.~Hong, and N.~Lee, ``Blind detection for {MIMO} systems with
  low-resolution {ADC}s using supervised learning,'' in \emph{2017 IEEE
  International Conference on Communications (ICC)}, May 2017, pp. 1--6.

\bibitem{Shao18}
Z.~{Shao}, R.~C. {de Lamare}, and L.~T.~N. {Landau}, ``Iterative detection and
  decoding for large-scale multiple-antenna systems with 1-bit {ADC}s,''
  \emph{IEEE Wireless Communications Letters}, vol.~7, no.~3, pp. 476--479,
  Jun. 2018.

\bibitem{Jeon18}
Y.~{Jeon}, N.~{Lee}, S.~{Hong}, and R.~W. {Heath}, ``A low complexity {ML}
  detection for uplink massive {MIMO} systems with one-bit {ADC}s,'' in
  \emph{2018 IEEE 87th Vehicular Technology Conference (VTC Spring)}, Jun.
  2018, pp. 1--5.

\bibitem{Cho19}
Y.~{Cho} and S.~{Hong}, ``One-bit successive-cancellation soft-output ({OSS})
  detector for uplink {MU-MIMO} systems with one-bit {ADC}s,'' \emph{IEEE
  Access}, vol.~7, pp. 27\,172--27\,182, 2019.

\bibitem{Choi16}
J.~Choi, J.~Mo, and R.~W. Heath, ``{N}ear {m}aximum-{l}ikelihood {d}etector and
  {c}hannel {e}stimator for {u}plink {m}ultiuser {m}assive {MIMO} {s}ystems
  {w}ith {o}ne-{b}it {ADC}s,'' \emph{IEEE Transactions on Communications},
  vol.~64, no.~5, pp. 2005--2018, May 2016.

\bibitem{Wen16}
C.~K. Wen, C.~J. Wang, S.~Jin, K.~K. Wong, and P.~Ting, ``Bayes-optimal joint
  channel-and-data estimation for massive {MIMO} with low-precision {ADC}s,''
  \emph{IEEE Transactions on Signal Processing}, vol.~64, no.~10, pp.
  2541--2556, May 2016.

\bibitem{Jmo17}
J.~Mo, P.~Schniter, and R.~W. Heath, ``Channel estimation in broadband
  millimeter wave {MIMO} systems with few-bit {ADC}s,'' \emph{IEEE Transactions
  on Signal Processing}, vol.~66, no.~5, pp. 1141--1154, Mar. 2018.

\bibitem{Jmo171}
J.~Mo, A.~Alkhateeb, S.~Abu-Surra, and R.~W. Heath, ``Hybrid architectures with
  few-bit {ADC} receivers: Achievable rates and energy-rate tradeoffs,''
  \emph{IEEE Transactions on Wireless Communications}, vol.~16, no.~4, pp.
  2274--2287, Apr. 2017.

\bibitem{Shao19}
Z.~{Shao}, L.~T.~N. {Landau}, and R.~C. {de Lamare}, ``Channel estimation using
  1-bit quantization and oversampling for large-scale multiple-antenna
  systems,'' in \emph{ICASSP 2019 - 2019 IEEE International Conference on
  Acoustics, Speech and Signal Processing (ICASSP)}, May 2019, pp. 4669--4673.

\bibitem{Bussgang52}
J.~J. Bussgang, ``Crosscorrelation functions of amplitude-distorted {G}aussian
  signals,'' \emph{MIT Res. Lab. Elec. Tech. Rep.}, vol. 216, pp. 1--14, 1952.

\bibitem{Kalman}
S.~M. Kay, \emph{Fundamentals of Statistical Signal Processing: Estimation
  Theory}, 1st~ed.\hskip 1em plus 0.5em minus 0.4em\relax New Jersey: Prentice
  Hall, 2000.

\bibitem{Li17}
Y.~Li, C.~Tao, G.~Seco-Granados, A.~Mezghani, A.~L. Swindlehurst, and L.~Liu,
  ``Channel estimation and performance analysis of one-bit massive {MIMO}
  systems,'' \emph{IEEE Transactions on Signal Processing}, vol.~65, no.~15,
  pp. 4075--4089, Aug. 2017.

\bibitem{Clerckx08}
B.~Clerckx, G.~Kim, and S.~Kim, ``Correlated fading in broadcast {MIMO}
  channels: curse or blessing?'' in \emph{2008 IEEE Global Telecommunications
  Conference}, Nov. 2008, pp. 1--5.

\bibitem{Choi141}
J.~{Choi}, D.~J. {Love}, and D.~R. {Brown}, ``Channel estimation techniques for
  quantized distributed reception in {MIMO} systems,'' in \emph{2014 48th
  Asilomar Conference on Signals, Systems and Computers}, Nov. 2014, pp.
  1066--1070.

\bibitem{arcsin}
G.~Jacovitti and A.~Neri, ``Estimation of the autocorrelation function of
  complex {G}aussian stationary processes by amplitude clipped signals,''
  \emph{IEEE Transactions on Information Theory}, vol.~40, no.~1, pp. 239--245,
  Jan. 1994.

\bibitem{Shariati14}
N.~Shariati, E.~Björnson, M.~Bengtsson, and M.~Debbah, ``Low-complexity
  polynomial channel estimation in large-scale {MIMO} with arbitrary
  statistics,'' \emph{IEEE Journal of Selected Topics in Signal Processing},
  vol.~8, no.~5, pp. 815--830, Oct. 2014.

\bibitem{Diggavi01}
S.~N. Diggavi and T.~M. Cover, ``The worst additive noise under a covariance
  constraint,'' \emph{IEEE Transactions on Information Theory}, vol.~47, no.~7,
  pp. 3072--3081, Nov. 2001.

\bibitem{Choi12}
J.~Choi, B.~Clerckx, N.~Lee, and G.~Kim, ``A new design of polar-cap
  differential codebook for temporally/spatially correlated {MISO} channels,''
  \emph{IEEE Transactions on Wireless Communications}, vol.~11, no.~2, pp.
  703--711, Feb. 2012.

\bibitem{Kotecha04}
J.~H. {Kotecha} and A.~M. {Sayeed}, ``Transmit signal design for optimal
  estimation of correlated {MIMO} channels,'' \emph{IEEE Transactions on Signal
  Processing}, vol.~52, no.~2, pp. 546--557, Feb. 2004.

\bibitem{Bjornson10}
E.~{Bjornson} and B.~{Ottersten}, ``A framework for training-based estimation
  in arbitrarily correlated {R}ician {MIMO} channels with {R}ician
  disturbance,'' \emph{IEEE Transactions on Signal Processing}, vol.~58, no.~3,
  pp. 1807--1820, Mar. 2010.

\bibitem{Choi142}
J.~{Choi}, D.~J. {Love}, and P.~{Bidigare}, ``Downlink training techniques for
  {FDD} massive {MIMO} systems: open-loop and closed-loop training with
  memory,'' \emph{IEEE Journal of Selected Topics in Signal Processing},
  vol.~8, no.~5, pp. 802--814, Oct. 2014.

\bibitem{Dither14}
U.~Gustavsson, C.~Sanchéz-Perez, T.~Eriksson, F.~Athley, G.~Durisi, P.~Landin,
  K.~Hausmair, C.~Fager, and L.~Svensson, ``On the impact of hardware
  impairments on massive {MIMO},'' in \emph{2014 IEEE Globecom Workshops (GC
  Wkshps)}, Dec. 2014, pp. 294--300.

\bibitem{Bibby74}
J.~Bibby, ``Axiomatisations of the average and a further generalisation of
  monotonic sequences,'' \emph{Glasgow Mathematical Journal}, vol.~15, no.~1,
  p. 63–65, 1974.

\end{thebibliography}

\end{document}